\documentclass[a4paper, amsfonts, amssymb, amsmath, reprint, showkeys, nofootinbib]{revtex4-1}
\usepackage[english]{babel}
\usepackage[utf8]{inputenc}
\usepackage[colorinlistoftodos, color=green!40, prependcaption]{todonotes}
\usepackage{amsthm}
\usepackage{mathtools}
\usepackage{physics}
\usepackage{xcolor}
\usepackage{graphicx}
\usepackage[left=19mm,right=16mm,top=20mm,columnsep=15pt]{geometry} 
\usepackage{adjustbox}
\usepackage{placeins}
\usepackage[T1]{fontenc}
\usepackage{lipsum}
\usepackage{csquotes}
\usepackage[pdftex, pdftitle={Article}, pdfauthor={Author}]{hyperref} 
\usepackage{siunitx}
\newcommand{\beginsupplement}{%
        \setcounter{table}{0}
        \renewcommand{\thetable}{S\arabic{table}}%
        \setcounter{figure}{0}
        \renewcommand{\thefigure}{S\arabic{figure}}%
        \renewcommand{\theHfigure}{Supplement.\thefigure}
     }

\usepackage{hyperref}
\newcommand{\ut}{\affiliation{Biomedical Photonic Imaging Group, Faculty of Science and Technology, University of Twente, P.O. Box 217, 7500 AE Enschede, The Netherlands}}

\makeatletter
\newcommand{\printfnsymbol}[2]{%
  \textsuperscript{\@fnsymbol{#1}}%
}
\makeatother
\begin{document}
\title{Model-based wavefront shaping microscopy}

\author{Abhilash Thendiyammal$^{*,\dagger}$}
\ut

\author{Gerwin Osnabrugge$^\dagger$}
\ut

\author{Tom Knop}
\ut

\author{Ivo M. Vellekoop}
\ut

\date{\today} 
\begin{abstract}
Wavefront shaping is increasingly being used in modern microscopy to obtain distortion-free, high-resolution images deep inside inhomogeneous media.  Wavefront shaping methods typically rely on the presence of a \textquoteleft{guidestar'} in order to find the optimal wavefront to mitigate the scattering of light. However, this condition cannot be satisfied in most biomedical applications. Here, we introduce a novel, guidestar-free wavefront shaping method in which the optimal wavefront is \emph{computed} using a digital model of the sample. The refractive index model of the sample, that serves as the input for the computation, is constructed in-situ by the microscope itself. In a proof of principle imaging experiment, we demonstrate a large improvement in the two-photon fluorescence signal through a diffuse medium, outperforming the state-of-the-art wavefront shaping techniques by a factor of 21.
\end{abstract}

\maketitle

\def\thefootnote{*}\footnotetext{Corresponding author: a.thendiyammal@utwente.nl}
\def\thefootnote{$\dagger$}\footnotetext{These authors contributed equally to this work}

\noindent
Imaging deep inside biological tissues at high resolution is a long sought-after goal in biomedical research. Light scattering due to inhomogeneities in the refractive index makes this task extremely challenging as scattering prevents the formation of a sharp focus and therefore deteriorates the image. This problem can be overcome by shaping the wavefront of the incident light to counteract the scattering. Recent progress in wavefront shaping has enabled control over light propagation through turbid media and imaging with sub-wavelength resolution \cite{kubby2019wavefront}.

In wavefront shaping, two main classes of approaches can be distinguished: feedback-based wavefront shaping \cite{mosk2012controlling} and optical phase conjugation \cite{Yaqoob2008}. Feedback-based wavefront shaping depends on the detection of the intensity feedback at a desired location to find the optimal wavefront maximizing that feedback signal. This signal can be obtained either by direct access through the sample \cite{vellekoop2007focusing} or using an embedded \textquoteleft{guidestar'} \cite{Horstmeyer2015,Vellekoop:08}. Although the improvement in the focus intensity can be remarkably high ($\approx$ 3 orders of magnitude \cite{vellekoop2015feedback}), this technique is limited to focusing onto the very guidestar used for feedback.

Alternatively, in optical phase conjugation the optimal wavefront is obtained from a single measurement of the scattered field propagating from a source located behind or inside the turbid medium \cite{Wang2015,Liu:17}. Subsequently, a focus is formed by playing back the conjugate of this field using a phase conjugate mirror. Rather than feedback from a guidestar, phase conjugation methods require a coherent light source to be present at the focus location. To form a coherent source, ultrasound can be focused to acoustically tag the scattered light at a desired location \cite{Xu2011,Wang2012}. Compared to light, ultrasound is not scattered as strongly in biological tissue, allowing the light to be tagged and focused at unprecedented depths of a few millimeters \cite{Wang2012}. However, the resolution of the focus depends on the ultrasound focus and is of the order of tens of micrometers.  

\begin{figure}[b]
\centering
\includegraphics[width=\linewidth]{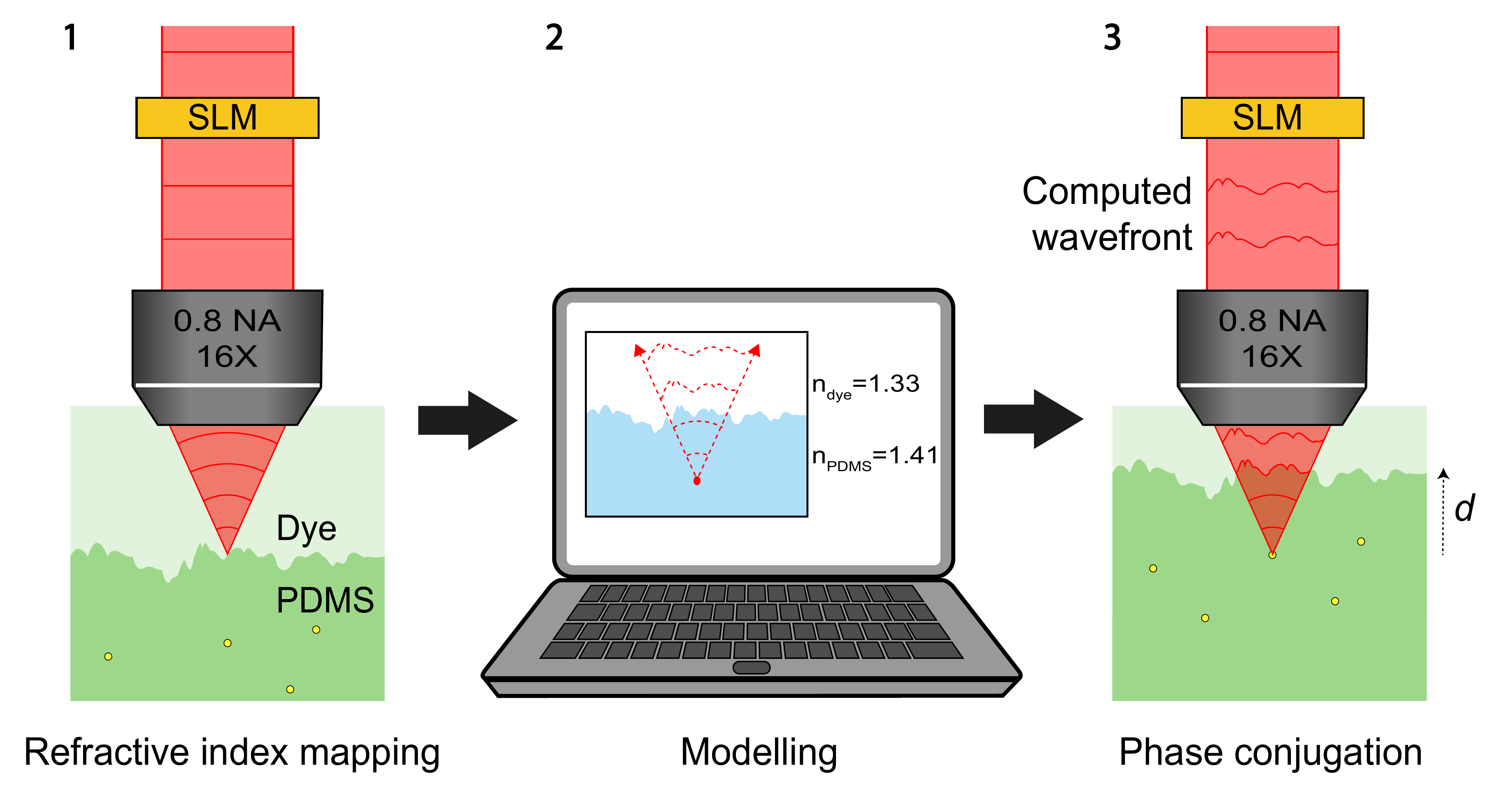}
\caption{\textbf{Principle of model-based wavefront shaping microscopy}. Step 1: We use two-photon fluorescence excitation microscopy to image the sample and generate a 3D refractive index map from the image data. Step 2: Light propagation inside the scattering sample is simulated to compute the wavefront required to focus at any desired location. Step 3: The computed wavefront is constructed with a spatial light modulator (SLM) in order to compensate for the scattering and to form a sharp focus.}
\label{fig:Concept}
\end{figure}

\begin{figure*}[t]
\centering
\includegraphics[width=\linewidth]{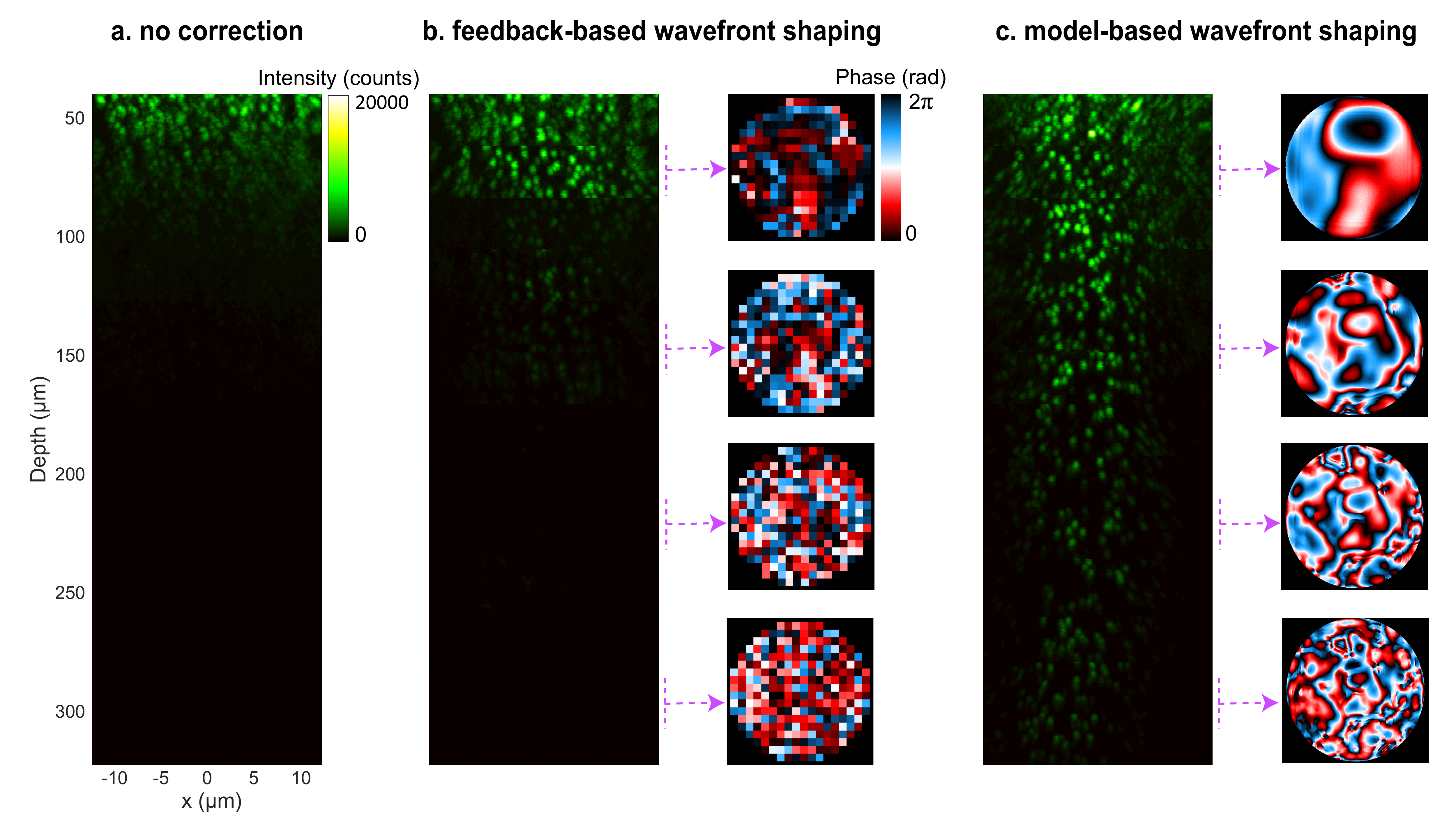}
\caption{\textbf{Scattering compensation using model-based wavefront shaping}. a, The maximum intensity projection of the 150 frames at the center of the 3D stack image acquired using conventional TPM.  The intensity from fluorescent beads decreases rapidly as a function of depth. The maximum intensity projection  of the TPM image after applying b, the correction wavefronts obtained from feedback-based method and c, the correction wavefronts obtained from model-based wavefront shaping. The wavefronts associated with four sub-stacks are also displayed. It is clear that model-based wavefront shaping works over the entire depth of interest, whereas the feedback-based method fails when noise dominates the feedback signal from the fluorescent beads.}
\label{fig:MIP}
\end{figure*}

Here, we introduce a \emph{third} class of  wavefront shaping methods, which we call model-based wavefront shaping. In this guidestar-free method, the optimal wavefront is computed numerically using a digital model of the sample. The microscopic refractive index model of the sample, that serves as the input for the calculations, is obtained from the image data itself.

The concept of model-based wavefront shaping is illustrated in Fig. \ref{fig:Concept}. In principle, we obtain the optimal wavefront by performing a virtual phase conjugation experiment. As a first step, we generate a refractive index distribution model by imaging the superficial layers of the scattering sample. In the second step, we place a \textquoteleft{virtual guidestar'} in our model and simulate the propagation of light from that point to outside the sample. Finally, the computed scattered field is phase conjugated and constructed with a spatial light modulator (SLM) to form a sharp focus. Once the refractive index model is generated, it is possible to compute wavefronts required to focus light anywhere inside the scattering medium.

As a proof of concept of this new method, we demonstrate enhanced imaging of 500 nm fluorescent beads through a single light-diffusing interface between polydimethylsiloxane (PDMS) and water. The fluorescent dye is added to the water to aid in visualising the interface. The sample is placed in a two-photon fluorescence excitation microscope (TPM) with an SLM conjugated to the back-pupil plane of the microscope objective. First, we use the microscope to acquire a 3D intensity image of the scattering surface. From this image, we reconstruct the 3D refractive index model of the PDMS-water interface (see supplementary information for the experimental setup, procedure for refractive index reconstruction, modelling, sample preparation, and data processing).

 We performed three imaging experiments to demonstrate the feasibility and robustness of our technique. In the first experiment, we use conventional TPM (with no correction for scattering) to image the beads. In the second experiment, we image the beads after applying the optimal wavefronts obtained from a state-of-the-art feedback-based wavefront shaping. In the third experiment, we image the beads after applying the optimal wavefronts computed using model-based wavefront shaping.
 
 Figure \ref{fig:MIP} illustrates our results where we compared the maximum intensity projections of the acquired TPM images computed along the y-axis. Figure \ref{fig:MIP}(a) shows the maximum intensity projection of the 3D stack acquired using conventional TPM. We have combined thirteen 3D sub-stacks to cover the depth (z-axis) range from \SI{42}{\micro m} to \SI{325}{\micro m} through the scattering layer. Each 3D sub-stack consists of 41 frames with a volume of 25.6x25.6x21.7 \si{\micro m^3}. It is clear from the Fig. \ref{fig:MIP}(a) that the intensity of the image  decreases rapidly as a function of distance from the scattering layer. 

 Figure \ref{fig:MIP}(b) shows the maximum intensity projection of the TPM image after applying the correction wavefronts obtained from a feedback-based method. Feedback-based optimisation has been carried out using a Hadamard algorithm \cite{Tao:17} with 256 input modes. For every 3D sub-stack, we found the optimal wavefront by optimising the feedback signal from a single fluorescent bead located at the center. To image each 3D sub-stack, we used a single wavefront correction. It is clear from Fig. \ref{fig:MIP}(b) that the intensities of the beads are higher than that in Fig. \ref{fig:MIP}(a). However, as the signal-to-noise ratio (SNR) decreases with depth, the feedback-based method fails to optimise the focus after about a depth of \SI{175}{\micro m}. It can also be seen that the intensity does not follow a monotonic variation over these depths due to the low SNR during the optimisation procedure. 

Figure \ref{fig:MIP}(c) shows the maximum intensity projection of the TPM image after applying the correction wavefronts obtained from model-based wavefront shaping. We used a beam propagation method (BPM) adopting the angular spectrum method \cite{goodman2005introduction}  for simulating the light propagation through the sample and computing the optimal wavefront  for phase conjugation. For every 3D sub-stack, the image intensity was enhanced by computing a single wavefront required to focus at the center position. Using a standard desktop PC, BPM simulations took less than 30 seconds to find the optimal wavefront. In contrast to feedback-based wavefront shaping, the fluorescent beads are visible all the way to the maximum depth of 325 \si{\micro\metre}, which is approximately twice the depth reached by feedback-based wavefront shaping. 

The optimised and computed wavefronts corresponding to four different sub-stacks are shown in the Figs. \ref{fig:MIP}(b) and \ref{fig:MIP}(c). The model-based wavefront shaping finds an accurate representation of the correction wavefront, which becomes more complex with increasing depth. This result is a clear improvement over feedback-based wavefront shaping, where the resolution of the correction is fixed by the algorithm, and the quality decreases with depth.
 
 \begin{figure}
\centering
\includegraphics[width=\linewidth]{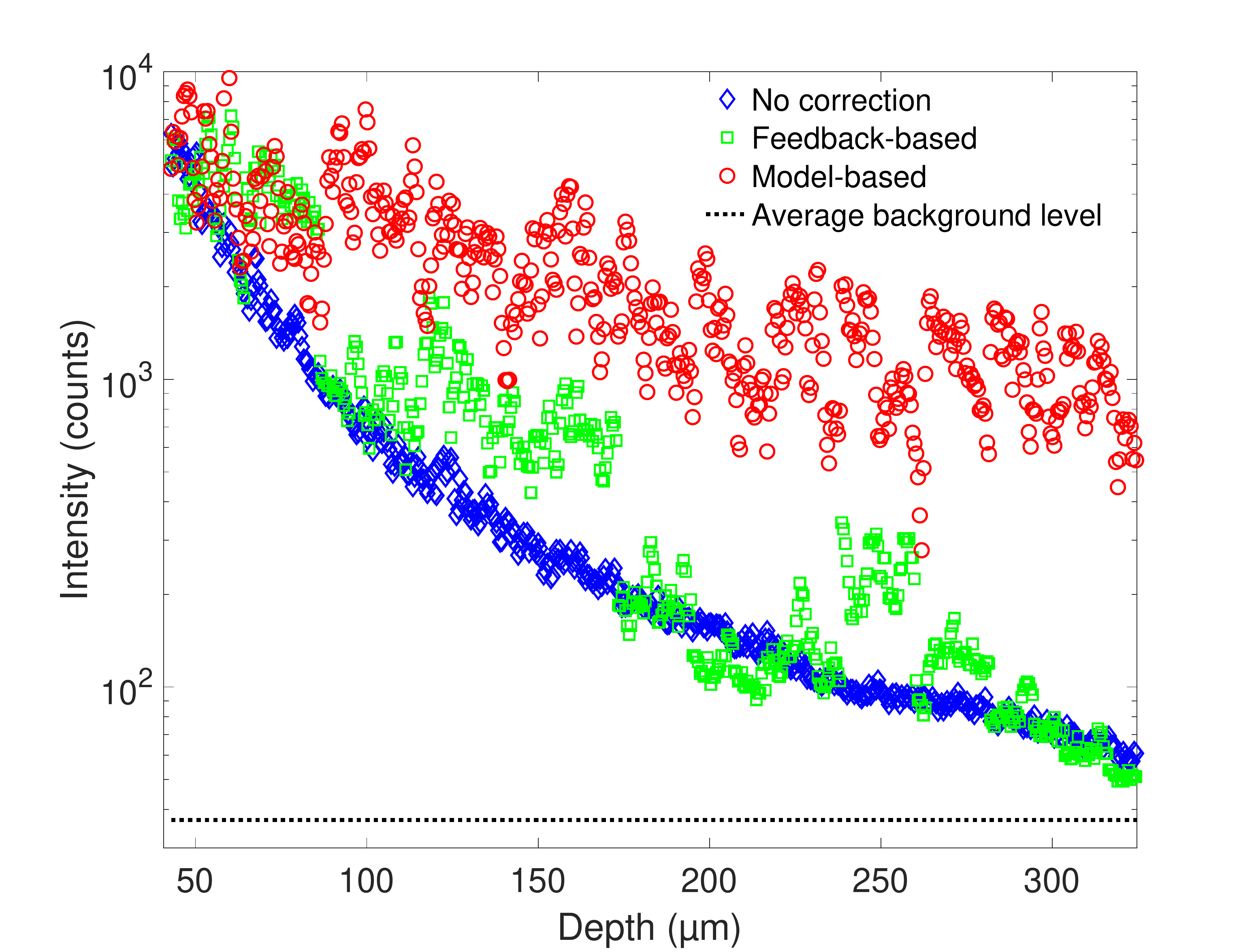}
\caption{\textbf{Fluorescent signal from 500 nm beads as a function of depth inside the PDMS diffuser before and after correction}. Open diamonds represent the intensity before applying correction wavefront on the SLM. Open squares and circles represent respectively the intensities after applying correction wavefronts obtained from feedback-based wavefront shaping and model-based wavefront shaping.
}
\label{fig:Graph}
\end{figure}
 
Figure \ref{fig:Graph} depicts the two-photon signal as a function of depth before and after compensating for the scattering. The average background level of the intensity is also plotted as a reference (dashed line). A larger area of the scattering surface is illuminated by the focusing beam as the imaging depth increases. Therefore, the uncorrected fluorescent signal (diamonds) drops rapidly as a function of depth. The feedback-based method (squares) successfully enhances the image intensity until about a depth of \SI{175}{\micro m} but fails to improve the focus at larger depths because of the drop in SNR. Model-based wavefront shaping works over the entire depth of the 3D image and shows a 21-fold increase in intensity (circles) at the deepest optimised point. It is to be noted that the intensity slowly decreases with depth even after correction. This may be due to the absorption of light by the fluorescent beads or by small inaccuracies in the modelling or in the experimental alignment.

In conclusion, this work introduces a new class of wavefront shaping methods combining TPM imaging and light propagation modelling to mitigate scattering in a robust way. The main advantage of our technique over other methods is that it does not require any guidestar for finding the optimal wavefront. Therefore, many practical limitations like SNR, number of optimised modes, etc. associated with the other techniques can be ignored. The primary step in our method is the generation of a refractive index model. This can be done directly from the microscope as we did here, or one may use techniques such as optical diffraction tomography, optical coherence tomography, ptychography, structured illumination microscopy, etc. \cite{Choi2007, Callewaert:17, Chowdhury:19, Chowdhury:17}. We envision that model-based wavefront shaping will play a key role for deep-tissue microscopy at depths where isolated guidestars are no longer visible.

\section*{Funding}
The research leading to these results has received funding from the European Research Council under the European Union's Horizon 2020 Programme / ERC Grant Agreement $\text{n}^\circ$ [678919].

\section*{ACKNOWLEDGEMENTS}
The authors would like to thank Tzu Lun Ohn for providing the protocol for the sample preparation. 
\bibliography{sample.bib}
\onecolumngrid
\newpage
\beginsupplement
\section*{\textbf{Supplementary Information A: The experimental setup}}
\begin{figure*}[!h]
\centering
\includegraphics[width=\linewidth]{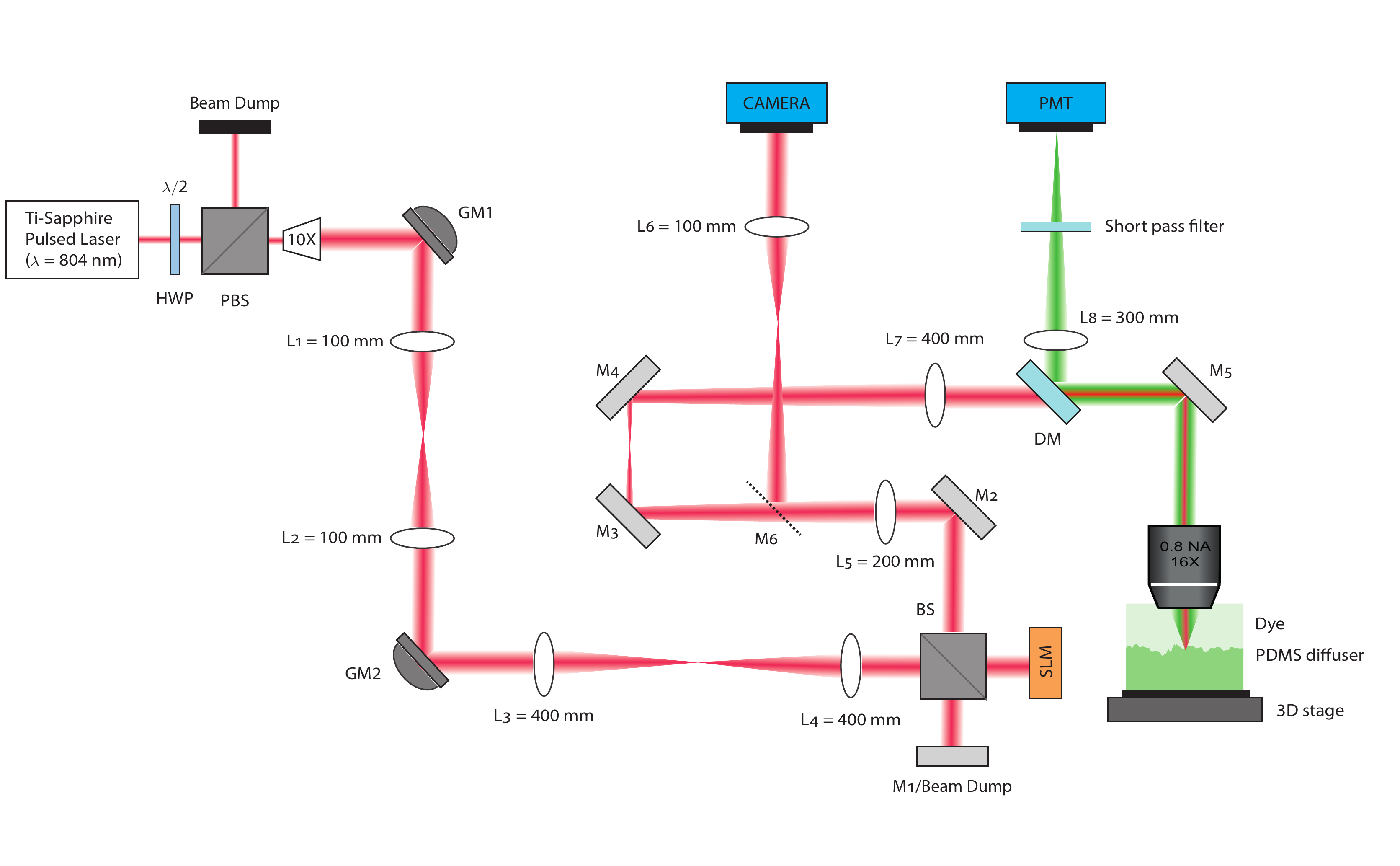}
\caption{\textbf{Experimental setup}. PBS, polarizing beam splitter, HWP, half-wave plate, GM, galvo mirror, L, lens, M, mirror, BS, 50/50 beam splitter, DM, Dichroic mirror, SLM, spatial light modulator.  M6 is a flip mirror used to facilitate the imaging of the SLM during initial calibration measurements.}
\label{fig:Setup}
\end{figure*}

\noindent
The setup used for two-photon fluorescence excitation microscopy (TPM) is illustrated in Fig. \ref{fig:Setup}. A titanium-sapphire laser (Spectra-Physics, Mai Tai) is used as the light source for two-photon excitation at a wavelength of 804 nm. The power and the polarization of the laser beam are controlled using a half-wave plate (HWP)  and a polarizing beam splitter (PBS). The laser beam is expanded and sent to two galvo mirrors (GM) which are used for scanning the beam. A spatial light modulator (SLM, LC, Meadowlark Optics, $1920\times1152$ pixels) is conjugated to the pupil plane of the objective lens (Nikon, CFI75 LWD 16x, numerical aperture of 0.8). A photomultiplier tube (PMT, Hamamatsu, H10770(P)A-40/-50) is used to detect the fluorescent light emitted by the sample. To collect only the fluorescent light, a dichroic mirror (Semrock, FF685-Di02-25$\times$36) and a short pass transmission filter (Semrock, FF01-680/SP-25) are used. By flipping the mirror M6, the light can be redirected to a camera (Basler, acA2000-165umNIR), which is is used to image the SLM for initial calibration measurements. The sample is placed on a 3D stage to facilitate initial alignment. Mirror M1 is used for a reference arm during the calibration measurements and is replaced with a beam dump during TPM imaging. A piezo scanning stage (PI, PD72Z2x/4x) is used to move the objective lens for depth scanning.

\section*{Supplementary Information B: Sample preparation} 
\begin{figure*}[!h]
\centering
\includegraphics[width=\linewidth/2]{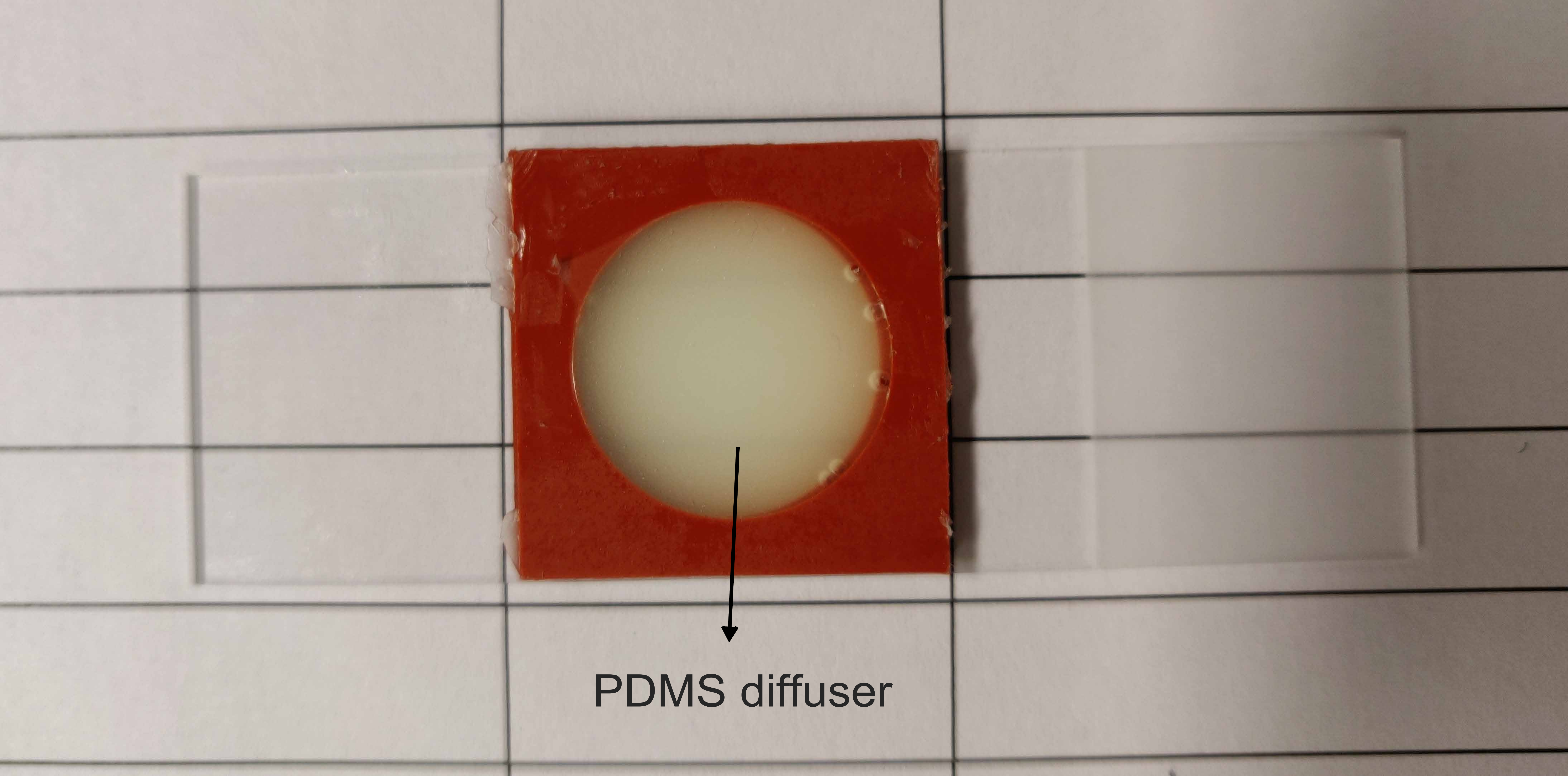}
\caption{\textbf{Sample}. PDMS diffuser dispersed with 500 nm fluorescent beads.}
\label{fig:Sample}
\end{figure*}

\noindent
We used the following protocol to make a polydimethylsiloxane   (PDMS) diffuser dispersed with fluorescent beads. First, we mixed the fluorescent beads (Fluoresbrite, plain YG, 500 nm microspheres) with Triton X-100 + Water + Ethanol (1:1:1) solution in 1:2 ratio in order to avoid cluster formation. The resulting solution is mixed with PDMS (DMS base + Curing agent in 10:1 ratio, Sylgrad  184,  Dow  Silicones)  in a 1:66 ratio. In order to disperse the microspheres uniformly, we grinded this mix for 20 minutes. After that we put the solution in a vacuum chamber and removed the air bubbles. The resulting solution is centrifuged at 2000 RPM for 5 minutes. The single diffusive layer of PDMS is formed by allowing this mix to cure on the surface of a ground glass diffuser (120 grit, custom-made) at \SI{50}{\degree C} for 2 hours. Fig. \ref{fig:Sample} shows an image of the sample.

\section*{Supplementary Information C: 3D Refractive index reconstruction}
\begin{figure*}[h!]
\centering
\includegraphics[width=\linewidth]{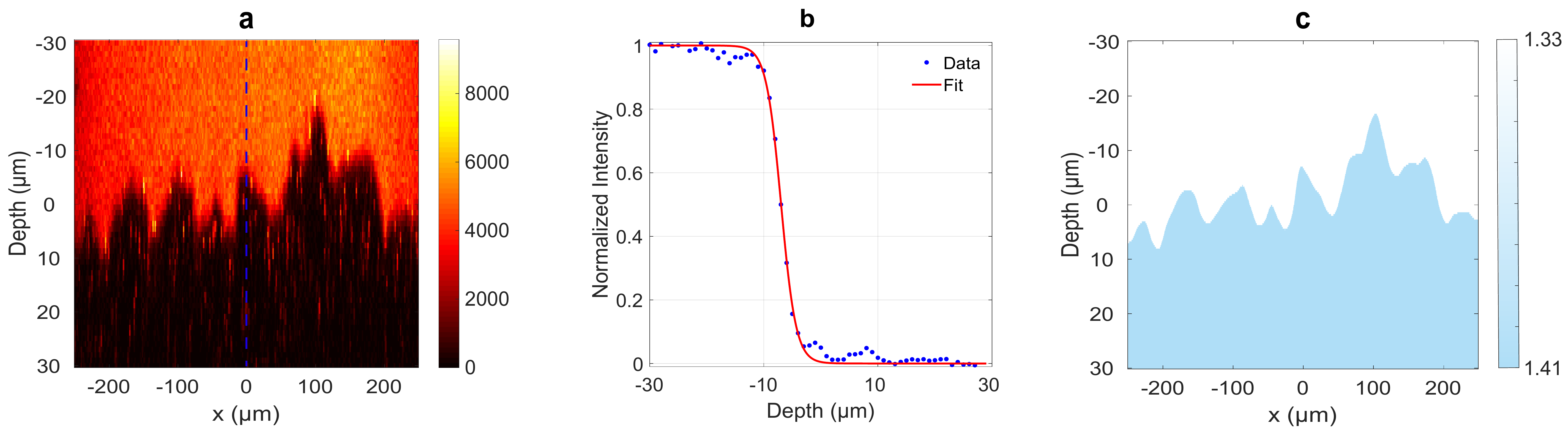}
\caption{\textbf{Refractive index reconstruction}. a, A 2D cross-section (middle frame) of the acquired 3D TPM image of the PDMS-water interface. b, An example fit using a sigmoid function. c, A 2D cross-section of the reconstructed refractive index distribution.}
\label{fig:RI}
\end{figure*}

\noindent
We use TPM to image the interface between the PDMS diffuser and water. In order to visualize the surface, a dye of fluorescein is added to the water (1 mg/mL, Sigma-Aldrich). We have acquired 60 TPM frames covering a volume of 500$\times$500$\times$60 \si{\micro m^3}. Fig. \ref{fig:RI}(a) shows a 2D cross-section of the 3D image of the sample. Two separate regions can be identified in the figure. The bright region corresponds to fluorescein dye, whereas the dark region with localised high intensity spots corresponds to PDMS dispersed with fluorescent beads. A nonlinear fitting procedure has been implemented to automatically detect the PDMS-water interface. We first apply a low pass filter to the frames to remove the high intensity spots. Next, for every position, sigmoid functions are fitted to the intensity data along the depth. For every fit, we computed the point of inflection, which is assumed to be the depth of PDMS-water interface. An example fit at the center of the frame along the blue dashed line is shown in Fig. \ref{fig:RI}(b). We then assign the refractive index values 1.33 and 1.41 to the regions of water and PDMS, respectively. A 2D cross-section of the reconstructed refractive index is shown in the Fig. \ref{fig:RI}(c).

\section*{Supplementary Information D: Modelling}
\begin{figure*}
\centering
\includegraphics[width=\linewidth*3/5]{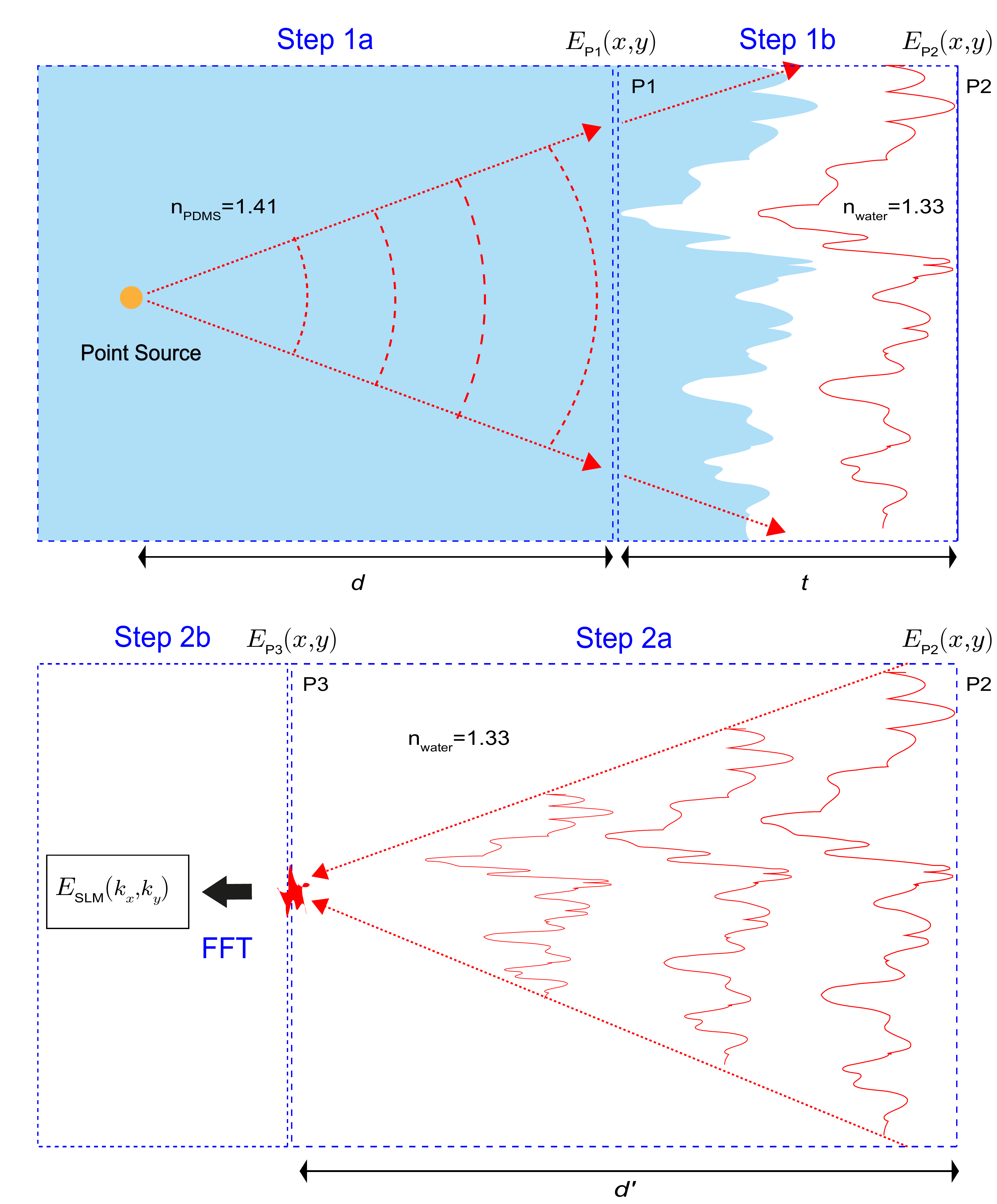}
\caption{\textbf{Implementation of the beam propagation method}. Step 1a, Generation of a spherical wavefront (at plane P1) from a point source located at a distance $d$ inside the PDMS medium. Step 1b, BPM is used to propagate this wavefront to plane P2 through the reconstructed refractive index of the PDMS-water interface. Step 2a, We propagate the scattered field at P2 back through the water to a plane P3. Step 2b, The field at P3 is propagated to the pupil plane of the microscope objective with a two-dimensional Fourier transform.}
\label{fig:BPM}
\end{figure*}

\noindent
We use a beam propagation method (BPM) adopting the angular spectrum method \cite{goodman2005introduction} to simulate the light propagation from a point source inside the scattering sample. In principle, we simulate the recording step of a phase conjugation experiment. As the SLM is conjugated to the pupil plane of the objective lens, ideally what is required is a light propagation simulation through the sample, objective lens and other components of the microscope setup to the position of the SLM. This procedure is computationally cumbersome, and therefore we neglect the aberrations introduced by the optical components in the setup. We followed a simple procedure consisting of two steps to find the correction wavefront. The two steps consists of four sub-steps which are indicated by the blue-dashed rectangular regions in the Fig. \ref{fig:BPM}.

\textbf{1}. As a first step, we propagate light from a point source inside the PDMS sample to outside the scattering surface. In step 1(a), we analytically generate a spherical wavefront from a point source located at a distance $\textit{d}$ inside the PDMS. A transmission function of the form $\exp \{i \frac{2\pi {n_{\text {PDMS }}}}{\lambda}[\sqrt{d^{2}+x^{2}+y^{2}}-|d|]\}$, is used to model the diverging beam. ${n_{\text {PDMS }}}$ is the refractive index of PDMS, $\lambda$ is the wavelength of the light, $\textit{x}$ and $\textit{y}$ are the spatial coordinates of the electric field, $E_{\text {P}{1}}(\textit{x},\textit{y})$ at plane P1. The opening angle of the point source is chosen to correspond to the numerical aperture of the microscope objective.

In step 1(b), we use BPM to propagate the field $E_{\text {P}{1}}(\textit{x},\textit{y})$ through the reconstructed refractive index distribution of the interface between the PDMS and water. For BPM, we first convert the refractive index distribution over a depth of \textit{t} = 60 \si{\micro\metre} into 180 layers of equally-spaced infinitely thin phase plates providing an approximation to the 3D distribution \cite{Yang:19}. The resulting computed field is $E_{\text {P}{2}}(\textit{x},\textit{y})$, which is located outside the sample at plane P2. 

\textbf{2}. In the second step, we propagate the scattered field  $E_{\text {P}{2}}(\textit{x},\textit{y})$ to the plane of the SLM. As our microscope objective is designed to be immersed in water, we first propagated the scattered field  $E_{\text {P}{2}}(\textit{x},\textit{y})$ back through
water over a distance $d^{\prime}=d \frac{n_{\text {water }}}{n_{\text {PDMS }}}$ to the plane P3. The computed field $E_{\text {P}{3}}(\textit{x},\textit{y})$ is then propagated to the pupil plane of the microscope objective with a two-dimensional Fourier transform to get  $E_{\text {SLM}}$($k_x$, $k_y$). Finally, the wavefront of the complex conjugate of $E_{\text {SLM}}$($k_x$, $k_y$) is generated with the SLM to form a sharp focus inside the sample.

\section*{Supplementary Information E: Calibration measurements for mapping SLM pixels to pupil plane}
\begin{figure*}[!h]
\centering
\includegraphics[width=\linewidth*3/4]{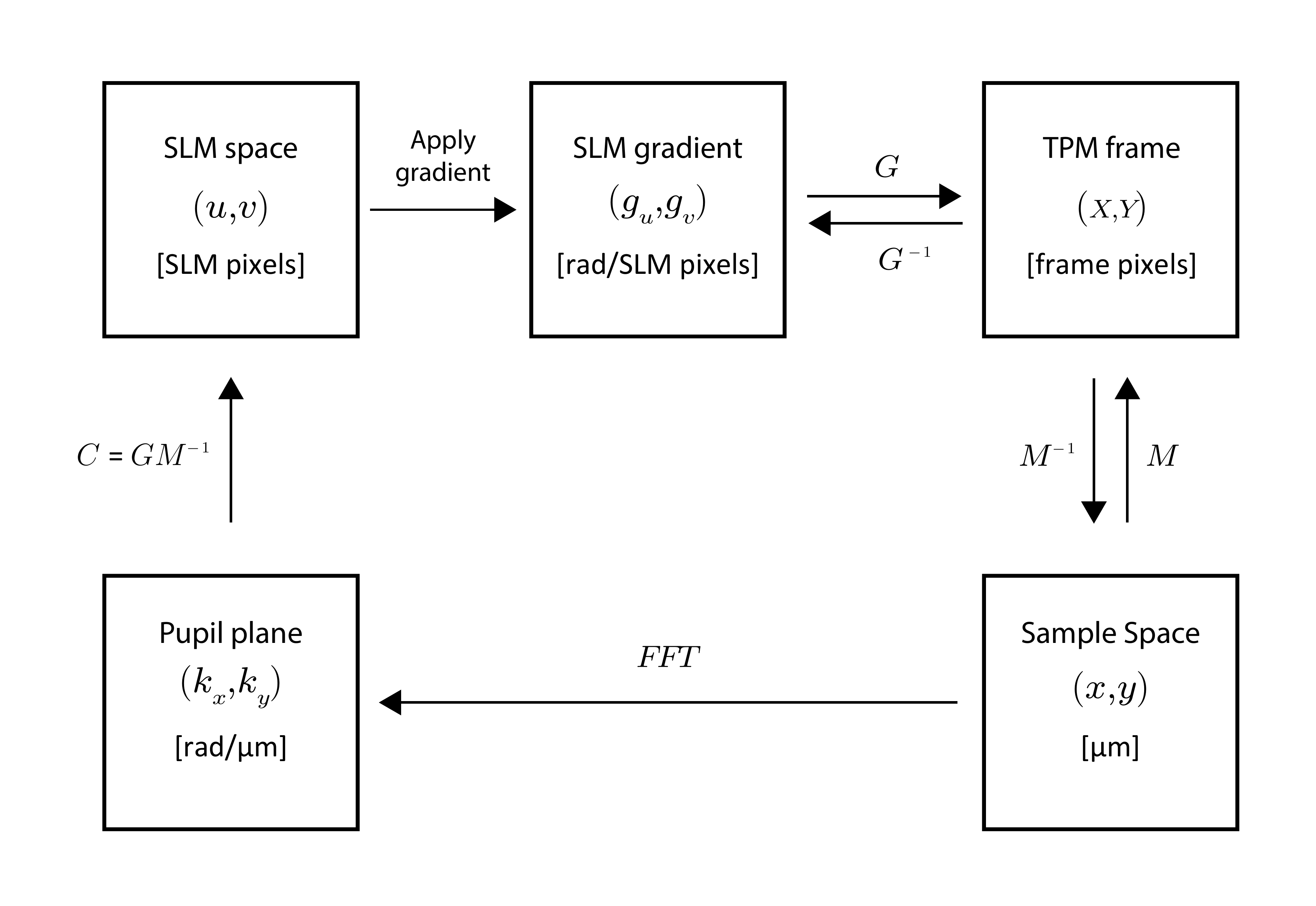}
\caption{\textbf{Block diagram showing the different geometrical spaces and the corresponding transformation matrices}.}
\label{fig:BlockDiagram}
\end{figure*}

\noindent
In order to successfully perform phase conjugation, the SLM pixels must be accurately mapped to the coordinates of the computer simulation. This is done by a sequence of calibration measurements. To go through the calibration procedure, let us first define the coordinates in the sample space as ($x$, $y$) in \SI{}{\micro m}, the coordinates in the TPM image space as ($X$, $Y$) in frame pixels, the coordinates in SLM space as ($u$, $v$) in SLM pixels and the spatial frequencies in the pupil plane as ($k_x$, $k_y$) in \SI{}{\radian}/\SI{}{\micro m}. Fig. \ref{fig:BlockDiagram} shows a block diagram with the different geometrical spaces and the transformation matrices connecting them. TPM imaging is carried out by scanning the angle of the incident beam using galvo mirrors. It is important to note that, as the galvo mirrors are scanning, the beam is standing still on the SLM and only the angle of incidence is changing. The steps in the calibration measurements are as follows,

\textbf{Step 1}:
We image the beam profile on the SLM with the camera (see Fig. \ref{fig:Setup}). From the camera image, we determine the center coordinates of the laser beam on the SLM.

\textbf{Step 2}: 
We calculate the field of view and the resolution of the TPM frame. We first calibrated the sample 2D stage (Zaber, ASR series microscope stage) with a resolution target (Thorlabs R1L3S6P). The two lateral stages in the Zaber are assumed to move orthogonally. In order to find the transformation matrix which converts pixel coordinates in the TPM frame to the coordinates in sample space, we displaced the sample with the Zaber stage. As the calibration sample, we used a 2D planar distribution of fluorescent beads made on top of a microscope slide. Initially an image of the sample is captured and saved as a reference frame. Then we captured two TPM frames with horizontal and vertical displacements using the lateral stages. We calculated the cross-correlations of the displaced frames with the initial reference frame. The peak positions in the cross-correlations are used to calculate the shifts in the frame pixels. In mathematical form, a transformation matrix ${ M}$ can be obtained by inverting the following relation:
\begin{equation}
\centering
\left( \begin{array} { l l } { \Delta X _ {1 } } & { \Delta X_ { 2 } } \\ { \Delta Y _ { 1 } } & { \Delta Y _ { 2 } } \end{array} \right)= M \left( \begin{array} { l l } {\Delta x} & { 0 } \\ { 0 } & {\Delta y} \end{array} \right),
\end{equation}
where ${ \Delta X_ { 1 } }, {\Delta Y _ { 1 } }, { \Delta X_ { 2 } }$ and ${ \Delta Y_ { 2 } }$ are image shifts (in frame pixels) observed in the TPM frame. $ \Delta x$ and $ \Delta y$ are the horizontal and vertical displacements (in \SI{}{\micro m}) applied on Zaber stage. The unit of transformation matrix ${ M}$ is [frame pixels/\SI{}{\micro m}].

\textbf{Step 3}: 
The next step is to find the transformation matrix that relates the SLM coordinates to the sample space coordinates. Initially, we captured the reference frame of the planar beads. Then we captured two TPM frames after applying horizontal and vertical shift to the frame by adding gradients along $u$-axis and $v$-axis on the SLM. We calculated the cross-correlations of the displaced frames with the initial reference frames. The peak positions in the cross-correlations are used to calculate the shifts in the frame pixels. The transformation matrix ${G}$ is given by,

\begin{equation}
\centering
\left( \begin{array} { l l } { \Delta X _ {1 } } & { \Delta X_ { 2 } } \\ { \Delta Y _ { 1 } } & { \Delta Y _ { 2 } } \end{array} \right) = G \left( \begin{array} { l l } g_ { u } & { 0 } \\ { 0 } &  g_ { v } \end{array} \right)
\end{equation}

where ${ \Delta X_ { 1 } }, {\Delta Y _ { 1 } }, { \Delta X_ { 2 } }$ and ${ \Delta Y_ { 2 } }$ are shifts (in frame pixels) in the TPM frame. ${ g_ { u } }$ and ${ g_ { v } }$ are the gradients (in \SI{}{\radian}/SLM pixels) applied on the SLM. The unit of ${ G}$ is [frame pixels$\times$SLM pixels/\SI{}{\radian}]. 

\textbf{Step 4}:
As SLM is conjugated to the pupil plane of the objective lens, the $k$-space coordinates in the simulation space have to be mapped to the SLM coordinates. This is done by combining the two transformation matrices ${M}$ and ${G}$ as follows,
\begin{equation}
\centering
    C = G M ^ { - 1 }
\end{equation}
The unit of ${C}$ is [SLM pixels$\times$\SI{}{\micro m}/\SI{}{\radian}]. 

By applying this conversion matrix on the $k$-space coordinates ($k_{x}$, $k_{y}$) in \SI{}{\radian}/\SI{}{\micro m} of the simulation space, we can find the SLM coordinates in pixels as follows,

\begin{equation}
\centering
    \left( \begin{array} { l } {u } \\ {v} \end{array} \right) = C \left( \begin{array} { l } { k _ { x } } \\ { k _ { y } } \end{array} \right)
\end{equation}

\centering
[SLM pixels] = [SLM pixels$\times$\SI{}{\micro m}/\SI{}{\radian}] $\times$ [\SI{}{\radian}/\SI{}{\micro m}] 

\end{document}